\documentclass[,nofootinbib,amsmath,amssym,preprint,aps]{revtex4}
\usepackage{hyperref,slashed,color}
\usepackage{graphicx,subfig}
\begin{document}
\preprint{TUHEP-TH-09167}
 \draft
\title{Electroweak Chiral Lagrangian for a Hypercharge-universal Topcolor Model}

\author{Jun-Yi Lang$^1$, Shao-Zhou Jiang$^1$,  Qing Wang$^{1,2}$\footnote{Corresponding author at:
Department of Physics, Tsinghua University, Beijing 100084,
P.R.China\\ {\it Email
address}:~wangq@mail.tsinghua.edu.cn(Q.Wang).}}

\address{$^1$Department of Physics,Tsinghua University,Beijing 100084,P.R.China \\
    $^2$Center for High Energy Physics, Tsinghua University, Beijing 100084, P.R.china}

\date{Jan 24, 2009}

\begin{abstract}
Electroweak chiral Lagrangian for a hypercharge-universal topcolor
model is investigated. We find that the assignments of universal
hypercharge improve the results obtained previously from K.Lane's
prototype natural TC2 model by allowing a larger $Z'$ mass resulting
in a very small T parameter and the S parameter is still around the
order of $+1$.

\end{abstract}
\maketitle


\vspace{1cm}
Topcolor-assisted technicolor (TC2) is a class of new physics models
which combines technicolor and topcolor together to realize
 the electroweak symmetry breaking dynamically. In these theories, a
 technicolor condensate provides the masses to the weak vector bosons
 and an extended technicolor (ETC) sector gives masses to the light
 quarks and leptons, and a bottom-quark-sized mass to the top. The
 majority of the top-quark mass is due to the formation of a top-quark
 condensate through the dynamics of an extended color gauge sector.
  The typical gauge group of the TC2 models is
 \begin{eqnarray}
 SU(N)_{TC}\otimes SU(3)_1\otimes SU(3)_2\otimes SU(2)_L\otimes U(1)_{Y_1}\otimes U(1)_{Y_2}
 \end{eqnarray}
, in which the extended color and hypercharge groups $SU(3)_1\otimes
SU(3)_2\otimes U(1)_{Y_1}\otimes U(1)_{Y_2}$ spontaneously break to
their diagonal subgroup $SU(3)_C\otimes U(1)_Y$ at a few TeVs and
the remaining electroweak groups $SU(2)_L\otimes U(1)_Y$
spontaneously break to their electromagnetic subgroup
$U(1)_{\mathrm{em}}$ at the electroweak scale due to a combination
of a top-quark condensate and a technifermion condensate. In the
original TC2 model \cite{Hill95,Lane95}, the extended hyper-charge
sector $U(1)_{Y_1}\otimes U(1)_{Y_2}$ is usually arranged
nonuniversal in flavor to ensure that the bottom-quark and other
light quarks and leptons do not condensate. Recently a new type of
TC2 model with a flavor-universal extended hyper-charge sector is
proposed in Ref.\cite{Sekhar}, the authors there have examined
various experimental and theoretical constraints, finding that
precision electroweak measurements yield the strongest bounds on the
model and the goodness of fit to all available Z-pole and LEP2 data
for hypercharge-universal topcolor is comparable to that of the
standard model (SM). In contrast, TC2 models with a flavor
nonuniversal hypercharge sector are markedly disfavored by the data.
The similar result on the nonuniversal hypercharge TC2 models is
also obtained from our works \cite{HongHao08, JunYi09}, where we
have computed the coefficients of the bosonic part of electroweak
chiral Lagrangian (EWCL) up to the order $p^4$ and found an upper
bound for the mass of flavor nonuniversal $Z'$ boson. For Hill's
schematic TC2 model \cite{Hill95}, $Z'$ mass $M_{Z'}$ is a few TeVs
and the S parameter can be either positive or negative depending on
whether the $M_{Z'}$ is large or small \cite{HongHao08}. While for
K.Lane's prototype natural TC2 model \cite{Lane95}, $M_{Z'}$ must be
smaller than 400GeV and the S parameter is around order of $+1$
\cite{JunYi09}. Since Ref.\cite{Sekhar} already shows explicitly the
experiment fit of the TC2 model due to the changes from the
 nonuniversal to the universal assignments for hypercharge sector, it is worthwhile
to apply our
 formulation developed in Ref.\cite{HongHao08} to the flavor-universal hypercharge topcolor model
 proposed in Ref.\cite{Sekhar} to examine the improvements from an alternative
 point of view.  Our formulation  offers an upper
 bound on nonuniversal $Z'$ mass previously, while Ref.\cite{Sekhar} gives a lower bound
of universal $Z'$ mass of roughly 2TeV. We expect that applying our
formulation to flavor-universal hypercharge topcolor model produces
an upper bound on universal $Z'$ mass which will compensate the
lower bound for the mass of universal $Z'$ boson obtained from
Ref.\cite{Sekhar}. In
 fact, from EWCL point of view, except technicolor and $Z'$ contributions,
 there are many other different sources to influence EWCL
 coefficients. In Ref.\cite{JunYi09}, we have made efforts to investigate
  the effective four-fermion interactions induced by extended technicolor
 (ETC). We find that their effects are small and we further point out that the walking technicolor (WTC)
 effects are worth future investigation.
 Considering that the authors in Ref.\cite{Sekhar} assume that WTC
 effects do not generate large precision electroweak corrections,
 up to present stage, we ignore WTC effects in this work.

In this paper, we are mainly interested in the effects from
flavor-universal hypercharge sector, to reduce the computations and
to be convenient for comparison with flavor-nonuniversal hypercharge
model, we base our calculations on the K.Lane's prototype natural
TC2 model \cite{Lane95} discussed in Ref.\cite{JunYi09}, but change
its hypercharge assignments to that given in Ref.\cite{Sekhar}. The
gauge charges are shown as Table I.

\begin{table}[h]
\small{{\bf TABLE I}.~Gauge charge assignments of techniquarks for
hypercharge universal TC2 model discussed in present paper. These
techniquarks are $SU(3)_1\otimes SU(3)_2$ singlets. }

\renewcommand{\arraystretch}{2}
\begin{tabular}{*{10}{c}}
field &~$T_L^l$~&~$U_R^l$~&~$D_R^l$~&~$T_L^t$~&~$U_R^t$~&~$D_R^t$~&~$T_L^b$~&~$U_R^b$~&~$D_R^b$\\
\hline {\footnotesize$SU(N)$} &N&N&N&N&N&N&N&N&N\\
{\footnotesize$SU(2)_L$} &2&1&1&2&1&1&2&1&1\\
{\footnotesize$U(1)_{\mbox{\tiny$Y_1$}}$}
&0&$\frac{1}{2}$&-$\frac{1}{2}$&0&$\frac{1}{2}$&-$\frac{1}{2}$&0&$\frac{1}{2}$&$-\frac{1}{2}$\\
{\footnotesize$U(1)_{\mbox{\tiny$Y_2$}}$}&0&0&0&0&0&0&0&0&0 \\
\hline
\end{tabular}
\end{table}
In later numerical computations, technicolor group representation
will be taken to be $N=3$.

 The action of the symmetry breaking sector is
\begin{eqnarray}
&&S_{\rm SBS}[G_{\mu}^\alpha,A_{1\mu}^A,A_{2\mu}^A,
W_\mu^a,B_{1\mu},B_{2\mu},\bar{T}^l,T^l,\bar{T}^t,T^t,\bar{T}^b,T^b]\nonumber\\
&&=\int d^4x({\cal L}_\mathrm{gauge}+{\cal
L}_\mathrm{techniquark}+\mathcal{L}_\mathrm{breaking}+\mathcal{L}_\mathrm{4T})\;,~~\label{SBSdef}
\end{eqnarray}
with ${\cal L}_\mathrm{techniquark}$,
$\mathcal{L}_\mathrm{breaking}$  and $\mathcal{L}_\mathrm{4T}$ being
the same as those in Ref.\cite{JunYi09} and the modified
 techinquark Lagrangian with flavor-universal hypercharge is
\begin{eqnarray}
\mathcal{L}_\mathrm{techniquark}&=&\bar{T}^{l}(i\slashed{\partial}-g_{\rm
TC}t^\alpha
\slashed{G}^\alpha-g_{2}\frac{\tau^{a}}{2}\slashed{W}^{a}P_{L}-\frac{1}{2}q_{1}\slashed{B}_{1}\tau^{3}P_{R})T^{l}
+\bar{T}^{t}(i\slashed{\partial}-g_{\rm TC}t^\alpha
\slashed{G}^\alpha-g_{2}\frac{\tau^{a}}{2}\slashed{W}^{a}P_{L}\nonumber\\
&&-\frac{1}{2}q_{1}\slashed{B}_{1}\tau^{3}P_{R})T^{t}
+\bar{T}^{b}(i\slashed{\partial}-g_{\rm TC}t^\alpha
\slashed{G}^\alpha-g_{2}\frac{\tau^{a}}{2}\slashed{W}^{a}P_{L}-\frac{1}{2}q_{1}\slashed{B}_{1}\tau^{3}P_{R})T^{b}\;.
\label{Ltechniquark}
\end{eqnarray}
 Rotating hypercharge gauge fields $B_{1\mu}$ and
$B_{2\mu}$ as
\begin{eqnarray}
&&\begin{pmatrix}B_{1\mu} &
B_{2\mu}\end{pmatrix}=\begin{pmatrix}Z_\mu^\prime &
B_\mu\end{pmatrix}
\begin{pmatrix}\cos\theta^\prime & -\sin\theta^\prime\\ \sin\theta^\prime &
\cos\theta^\prime\end{pmatrix}\;,\hspace{1cm}g_1=q_1\sin\theta'
=q_2\cos\theta'\;.~~~~\label{B1B2-BZpri}
\end{eqnarray}
The techinquark Lagrangian  (\ref{Ltechniquark}) is then reduced to
\begin{eqnarray}
{\cal L}_\mathrm{techniquark} =\bar{\psi}(i\slashed{\partial}-g_{\rm
TC}t^\alpha
\slashed{G}^\alpha+\slashed{V}+\slashed{A}\gamma^{5})\psi\;,~~
\end{eqnarray}
where all three doublets techniquarks are arranged in one by six
matrix $\psi=(U^l,D^l,U^t,D^t,U^b,D^b)^T$ and
\begin{eqnarray}
V_\mu=(-\frac{1}{2}g_2\frac{\tau^a}{2}W_{\mu}^a
-\frac{1}{2}g_1\frac{\tau^3}{2}B_{\mu})\otimes\mathbf{I}+Z_{V\mu}\hspace{1cm}A_\mu=(\frac{1}{2}g_2\frac{\tau^a}{2}W_{\mu}^a
-\frac{1}{2}g_1\frac{\tau^3}{2}B_{\mu})\otimes\mathbf{I}+Z_{A\mu}\;,\label{VAdef}
\end{eqnarray}
with $\mathbf{I}=\mathrm{diag}(1,1,1)$,
$Z_{V\mu}=\mathrm{diag}(Z_{V\mu}^l,Z_{V\mu}^t,Z_{V\mu}^b)$,
$Z_{A\mu}=\mathrm{diag}(Z_{A\mu}^l,Z_{A\mu}^t,Z_{A\mu}^b)$ and
\begin{eqnarray}
Z_{V\mu}^{l}=Z_{V\mu}^{t}=Z_{V\mu}^{b}=Z_{A\mu}^{l}=Z_{A\mu}^{t}=Z_{A\mu}^{b}=-\frac{1}{4}g_{1}\cot\theta'
Z^{\prime}_{\mu}\tau^{3}
\end{eqnarray}
As done in Ref.\cite{JunYi09}, the EWCL for present model is
\begin{eqnarray}
\exp\bigg(iS_{\mathrm{EW}}[W_\mu^a,B_\mu]\bigg)&=&\int\mathcal{D}\bar{\psi}\mathcal{D}\psi
\mathcal{D}G_\mu^\alpha\mathcal{D}Z'_{\mu}e^{iS_{\mathrm{SBS}}[G_{\mu}^\alpha\!,0,0,
W_\mu^a\!,B_{1\mu}\!,B_{2\mu}\!,\bar{T}^l\!,T^l\!,\bar{T}^t\!,T^t\!,\bar{T}^b\!,T^b]}~~\label{strategy-TC20}\nonumber\\
&=&\mathcal{N}[W_\mu^a,B_\mu]\int\mathcal{D}\mu(U)\exp\bigg(iS_{\mathrm{eff}}[U,W_\mu^a,B_\mu]\bigg)\;,
\label{strategy-TC2}
\end{eqnarray}
where $U(x)$ is a dimensionless unitary unimodular matrix field in
EWCL, and ${\cal D}\mu(U)$ denotes the normalized functional
integration measure on $U$. The normalization factor
$\mathcal{N}[W_\mu^a,B_\mu]$ is determined through the requirement
that when the technicolor interactions are switched off,
$S_{\mathrm{eff}}[U,W_\mu^a,B_\mu]$ must vanish.

The following computation procedure is exactly the same as those
given in Ref.\cite{JunYi09}, in which we integrated out the
technigluons, the techniquarks and the colorons. We abbreviate the
detailed process and only write down the resulted action,
\begin{eqnarray}
\int{\cal D}G_\mu^\alpha{\cal D}\bar{\psi}{\cal D}\psi{\cal
D}Z^{\prime}_{\mu}e^{iS_\mathrm{SBS}\big|_{A^A_{1\mu}=A^A_{2\mu}=0}}
=\int\mathcal{D}\mu(U){\cal
D}Z^{\prime}_{\mu}e^{iS_\mathrm{Z'}[U,W_\mu^a,B_\mu,Z^{\prime}_{\mu}]}\;,
\end{eqnarray}
with
\begin{eqnarray}
S_\mathrm{Z'}[U,W_\mu^a,B_\mu,Z^{\prime}_{\mu}]&=&-i\mathrm{Tr}\log(i\slashed{\partial}+\slashed{V}+\slashed{A}\gamma^{5})
+\int
d^4x\bigg[-\frac{1}{4}W_{\mu\nu}^aW^{a,\mu\nu}-\frac{1}{4}B_{\mu\nu}B^{\mu\nu}-\frac{1}{4}Z^{\prime}_{\mu\nu}Z^{\prime\mu\nu}
\nonumber\\
&&+\frac{1}{2}M_0^2Z^{\prime}_{\mu}Z^{\prime\mu}
+3\mathrm{tr}_f\bigg((F_0^{1D}
)^2a^{2}-\mathcal{K}_1^{1D,\Sigma\neq0}(d_\mu
a^{\mu})^2-\mathcal{K}_2^{1D,\Sigma\neq0}(d_\mu a_{\nu }-d_\nu
a_{\mu }
)^2\nonumber\\
&&+\mathcal{K}_3^{1D,\Sigma\neq0}(a^{2})^2+\mathcal{K}_4^{1D,\Sigma\neq0}(a_{\mu
} a_{\nu })^2 -\mathcal{K}_{13}^{1D,\Sigma\neq0}V_{\mu\nu }V^{\mu\nu
}+i\mathcal{K}_{14}^{1D,\Sigma\neq0}a_{\mu } a_{\nu
} V^{\mu\nu }\bigg)\bigg]\nonumber\\
&&+\mathcal{O}(p^6)\;,\label{allone}
\end{eqnarray}
where $M_0$ is the bare mass of $Z'$ boson from spontaneously
breaking of $SU(3)_1\otimes SU(3)_2\otimes U(1)_{Y_1}\otimes
U(1)_{Y_2}\Rightarrow SU(3)_C\otimes U(1)_Y$, as in
Ref.\cite{JunYi09} its relation with vacuum expectation value
$\tilde{v}$ causing breaking is
$M_0^2=\frac{25}{36}\frac{g_1^2\tilde{v^2}}{\sin^2\!\theta'\cos^2\!\theta'}$.
 The coefficients $F_0^{1D}$, $\mathcal{K}_i^{1D,\Sigma\neq0}$ for
$i=1,2,3,4,13,14$ are strong interaction coefficients for one
doublet technicolor model which depend on techniquark self energy
and are already computed numerically in
Ref.\cite{HongHao08,JunYi09}. Further
\begin{eqnarray}
v_\mu&\equiv&-\frac{1}{2}(g_2\frac{\tau^a}{2}W_{\xi\mu}^a
+g_1\frac{\tau^3}{2}B_{\xi\mu})-\frac{1}{4}g_{1}\cot\theta'
Z^{\prime}_{\mu}\tau^{3}\;,\\
a_\mu&\equiv&\frac{1}{2}(g_2\frac{\tau^a}{2}W_{\xi\mu}^a
-g_1\frac{\tau^3}{2}B_{\xi\mu})-\frac{1}{4}g_{1}\cot\theta'
Z^{\prime}_{\mu}\tau^{3}\;,
\end{eqnarray}
in which $W_{\xi\mu}^a$ and $B_{\xi\mu}$ are rotated electroweak
gauge fields given in Eq.(26) and (27) in Ref.\cite{JunYi09} which
absorb Goldstone field $U$ into the definition of gauge fields.

We can further decompose (\ref{allone}) into
\begin{eqnarray}
S_\mathrm{Z'}[U,W_\mu^a,B_\mu,Z^{\prime}_{\mu}]=
\tilde{S}_\mathrm{Z'}[U,W_\mu^a,B_\mu,Z^{\prime}_{\mu}]+S_\mathrm{Z'}[U,W_\mu^a,B_\mu,0]\;,
\end{eqnarray}
where $\tilde{S}_\mathrm{Z'}[U,W_\mu^a,B_\mu,Z^{\prime}_{\mu}]$ is
the $Z'$ dependent part of
$S_\mathrm{eff}[U,W_\mu^a,B_\mu,Z^{\prime}_{\mu}]$. We find that the
$Z'$ independent part $S_\mathrm{Z'}[U,W_\mu^a,B_\mu,0]$ is just the
same as that given in Ref.\cite{JunYi09} which is three times of the
one-doublet technicolor model result given in Ref.\cite{HongHao08}.
Similar as Ref.\cite{JunYi09}
$\tilde{S}_\mathrm{Z'}[U,W_\mu^a,B_\mu,Z^{\prime}_{\mu}]$ has the
structure
\begin{eqnarray}
\tilde{S}_\mathrm{Z'}[U,W_\mu^a,B_\mu,Z^{\prime}_{\mu}]=\int
d^4x~[\frac{1}{2}Z'_{R,\mu}D_Z^{-1,\mu\nu}Z'_{R,\nu}
+Z_R^{\prime,\mu}J_{Z,\mu}+Z_R^2Z_{R,\mu}'J^{\mu}_{3Z}
+g_{4Z}\frac{g_1^4}{c_{Z'}^{ 4}}Z_R^{\prime,4}]\;,~~~~\label{SZ'}
\end{eqnarray}
where $D_Z^{-1,\mu\nu}=g^{\mu\nu}(\partial^2+M^2_{Z'})
-(1+\lambda_Z)\partial^{\mu}\partial^{\nu}+\Delta^{\mu\nu}_Z(X)$
 and to normalize $Z'$ field correctly, we introduce normalized field
$Z^{\prime}_{R,\mu}$ as
$Z^{\prime}_{\mu}=\frac{1}{c_{Z'}}Z^{\prime}_{R,\mu}$. Due to the
present universal assignment of hypercharge, parameters
 appeared in $\tilde{S}_\mathrm{Z'}[U,W_\mu^a,B_\mu,Z^{\prime}_{\mu}]$
 are different from those in Ref.\cite{JunYi09},
\begin{eqnarray}
c_{Z'}^2&=&1+3\mathcal{K}g_{1}^{2}\cot^2\theta'
+\frac{3}{2}\mathcal{K}_{2}^{1D,\Sigma\neq0}g_{1}^{2}\cot^2\theta'
+\frac{3}{2}\mathcal{K}_{13}^{1D,\Sigma\neq0}g_{1}^{2}\cot^2\theta'\;,\label{cZ'def}\\
M_{Z'}^2&=&\frac{1}{c_{Z'}^{2}}\{M_0^2+\frac{3g_1^2\cot^2\theta'}{4}(F_0^{1D}
)^2\}\;,\label{MZdef}\\
\lambda_Z&=&-\frac{3g_1^2\cot^2\theta'}{4c_{Z'}^{2}}\mathcal{K}_1^{1D,\Sigma\neq0}\;,\\
\Delta^{\mu\nu}_Z(X)&=&\frac{g_1^2\cot^2\theta'}{16c_{Z'}^{2}}\bigg[(-12\mathcal{K}_{1}^{1D,\Sigma\neq0}-3\mathcal{K}_{3}^{1D,\Sigma\neq0}+6\mathcal{K}_{13}^{1D,\Sigma\neq0}-3\mathcal{K}_{14}^{1D,\Sigma\neq0})
\mathrm{tr}[X_{\mu}\tau^{3}]\mathrm{tr}[X^{\nu}\tau^{3}]\nonumber\\
&&+(24\mathcal{K}_1^{1D,\Sigma\neq0}-6\mathcal{K}_4^{1D,\Sigma\neq0}-12\mathcal{K}_{13}^{1D,\Sigma\neq0}+6\mathcal{K}_{14}^{1D,\Sigma\neq0})
\mathrm{tr}[X_{\mu}X^{\nu}]\nonumber\\
&&+g^{\mu\nu}(-3\mathcal{K}_3^{1D,\Sigma\neq0}+3\mathcal{K}_4^{1D,\Sigma\neq0}+12\mathcal{K}_{13}^{1D,\Sigma\neq0}-6\mathcal{K}_{14}^{1D,\Sigma\neq0})
\mathrm{tr}[X_{k}X^{k}]\notag\\
&&+g^{\mu\nu}(-3\mathcal{K}_4^{1D,\Sigma\neq0}-6\mathcal{K}_{13}^{1D,\Sigma\neq0}+3\mathcal{K}_{14}^{1D,\Sigma\neq0})
\mathrm{tr}[X_{k}\tau^{3}]\mathrm{tr}[X^{k}\tau^{3}]\bigg]\;,\\
J_Z^\mu&=&J_{Z0}^\mu+\frac{g_1^{2}\gamma}{c_{Z'}}\partial^{\nu}B_{\mu\nu}+\tilde{J}_Z^\mu\;,\label{JZdef}\\
 J_{Z0\mu}&=&\frac{3g_1\cot\theta'}{4c_{Z'}}i(F_0^{1D} )^2
\mathrm{tr}[X_{\mu}\tau^{3}]\;,\\
\gamma&=&-3\mathcal{K}\cot\theta'-\frac{3}{2}(\mathcal{K}_2^{1D,\Sigma\neq0}+\mathcal{K}_{13}^{1D,\Sigma\neq0})\cot\theta'\;,
\label{gammaDef}\\
\tilde{J}_{Z}^\mu&=&-\frac{g_1\cot\theta'}{4c_{Z'}}\bigg[\mathcal{K}_1^{1D,\Sigma\neq0}\{3i\mathrm{tr}[U^{\dag}(D^{\nu}D_{\nu}U)U^{\dag}D^{\mu}U\tau^{3}]
-3i\mathrm{tr}[U^{\dag}(D^{\nu}D_{\nu}U)\tau^{3}U^{\dag}D^{\mu}U]\nonumber\\
&&-3i\partial^{\mu}\mathrm{tr}[U^{\dag}(D^{\nu}D_{\nu}U)\tau^{3}]\}
+(-6\mathcal{K}_2^{1D,\Sigma\neq0}+6\mathcal{K}_{13}^{1D,\Sigma\neq0})\partial_{\nu}\mathrm{tr}[\overline{W}^{\mu\nu}\tau^{3}]\nonumber\\
&&+(\frac{3i}{4}\mathcal{K}_3^{1D,\Sigma\neq0}-\frac{3i}{4}\mathcal{K}_4^{1D,\Sigma\neq0}-3\mathcal{K}_{13}^{1D,\Sigma\neq0}+\frac{3i}{2}\mathcal{K}_{14}^{1D,\Sigma\neq0})
\mathrm{tr}[X^{\nu}X_{\nu}]\mathrm{tr}[X^{\mu}\tau^{3}]\nonumber\\
&&+(\frac{3i}{2}\mathcal{K}_4^{1D,\Sigma\neq0}+3\mathcal{K}_{13}^{1D,\Sigma\neq0}-\frac{3i}{2}\mathcal{K}_{14}^{1D,\Sigma\neq0})\mathrm{tr}[X^{\mu}X_{\nu}]\mathrm{tr}[X^{\nu}\tau^{3}]\nonumber\\
&&+(-3\mathcal{K}_{13}^{1D,\Sigma\neq0}+\frac{3}{4}\mathcal{K}_{14}^{1D,\Sigma\neq0})
\mathrm{tr}[\overline{W}^{\mu\nu
}(X_{\nu}\tau^{3}-\tau^{3}X_{\nu})]\nonumber\\
&&+(6i\mathcal{K}_{13}^{1D,\Sigma\neq0}-\frac{3}{2}i\mathcal{K}_{14}^{1D,\Sigma\neq0})
\partial_\nu\mathrm{tr}[X^{\mu}X^\nu\tau^{3}]\bigg]\;,\\
g_{4Z}&=&(\mathcal{K}_3^{1D,\Sigma\neq0}+\mathcal{K}_4^{1D,\Sigma\neq0})\frac{3\cot^4\theta'}{128}\;,\label{g4Zdef}\\
J_{3Z}^\mu&=&\frac{3ig_1^3\cot^3\theta'}{32c_{Z'}^{3}}(\mathcal{K}_3^{1D,\Sigma\neq0}+\mathcal{K}_4^{1D,\Sigma\neq0})
\mathrm{tr}[X^{\mu}\tau_{3}]\;,
\end{eqnarray}
where
\begin{eqnarray}
&&\hspace{-0.5cm}\mathcal{K}=-\frac{1}{48\pi^2}\left(\log\frac{\kappa^2}{\Lambda^2}+\gamma\right)\hspace{1cm}\Lambda,\kappa\mbox{:
ultraviolet and infrared cutoffs}\;.\label{kappaDef}
\end{eqnarray}
 With similar
procedure of Ref.\cite{JunYi09} to integrate out the $Z'$ field, we
find that $S_{\mathrm{eff}}[U,W_\mu^a,B_\mu]$ defined in
(\ref{strategy-TC2}) has exactly the standard structure of EWCL
given by Ref.\cite{EWCL}, from which we can read out coefficients up
to order of $p^4$ as follows,
\begin{eqnarray}
f^2&=&3(F_0^{1D})^2\;,\hspace{2cm}
\beta_1=\frac{3(F_0^{1D})^2g_1^2\cot^2\theta'}{8M^2_0+6(F_0^{1D})^2g_1^2\cot^2\theta'}\;,\label{f2beta1}\\
\alpha_{1}&=&3L_{10}^{1D}+\frac{3(F_0^{1D}
)^{2}}{2M_{Z'}^{2}}\beta_1+2\beta_1\tan\theta'\gamma-6\beta_1L_{10}^{1D}\;,\nonumber\\
\alpha_{2}&=&-\frac{3}{2}L_9^{1D}+\frac{3(F_0^{1D})^2}{2M_{Z'}^2}\beta_1+2\beta_1\tan\theta'\gamma+3\beta_1L_9^{1D}\;,\nonumber\\
\alpha_{3}&=&(-\frac{3}{2}+3\beta_1)L_9^{1D}\;,\nonumber\\
\alpha_{4}&=&3L_2^{1D}+6\beta_1L_9^{1D}+\frac{3(F_0^{1D})^2}{2M_{Z'}^2}\beta_1\;,\nonumber\\
\alpha_{5}&=&\frac{3}{2}L_3^{1D}+3L_1^{1D}-\frac{3(F_0^{1D})^2}{2M_{Z'}^2}\beta_1-6\beta_1L_9^{1D}\;,\nonumber\\
\alpha_{6}&=&-\frac{3(F_0^{1D})^2}{2M_{Z'}^2}\beta_1+24\beta_1^{2}L_1^{1D}-6\beta_1(4L_1^{1D}+L_9^{1D})\;,\label{alpha}\\
\alpha_{7}&=&\frac{3(F_0^{1D})^2}{2M_{Z'}^2}\beta_1+6\beta_1^2(L_3^{1D}+2L_1^{1D})-2\beta_1(3L_3^{1D}+6L_1^{1D}-3L_9^{1D})\;,\nonumber\\
\alpha_{8}&=&-\frac{3(F_0^{1D})^2}{2M_{Z'}^2}\beta_1+12\beta_1L_{10}^{1D}\;,\nonumber\\
\alpha_{9}&=&-\frac{3(F_0^{1D})^2}{2M_{Z'}^2}\beta_1+6\beta_1(L_{10}^{1D}-L_9^{1D})\;,\nonumber\\
\alpha_{10}&=&-4\beta_1^2(-18L_1^{1D}-3L_3^{1D})+32\beta_1^4\cot^4\theta'g_{4Z}-\frac{3}{2}\beta_1^3(96L_1^{1D}+16L_3^{1D})\;,\nonumber\\
\alpha_{11}&=&\alpha_{12}=\alpha_{13}=\alpha_{14}=0\;,\nonumber
\end{eqnarray}
where $L_i^{1D}$ for $i=1,3,9,10$ are EWCL coefficients for one
doublet technicolor model discussed in Ref.\cite{HongHao08}.

\begin{figure}[t]
\caption{The ratio of infrared cutoff and ultraviolet cutoff
$\kappa/\Lambda$ as functions of the $T$ parameter and $Z'$ mass in
unit of TeV. } \label{fig-cutoff1}
\hspace*{-4.5cm}\begin{minipage}[t]{5cm}
    \includegraphics[scale=0.6]{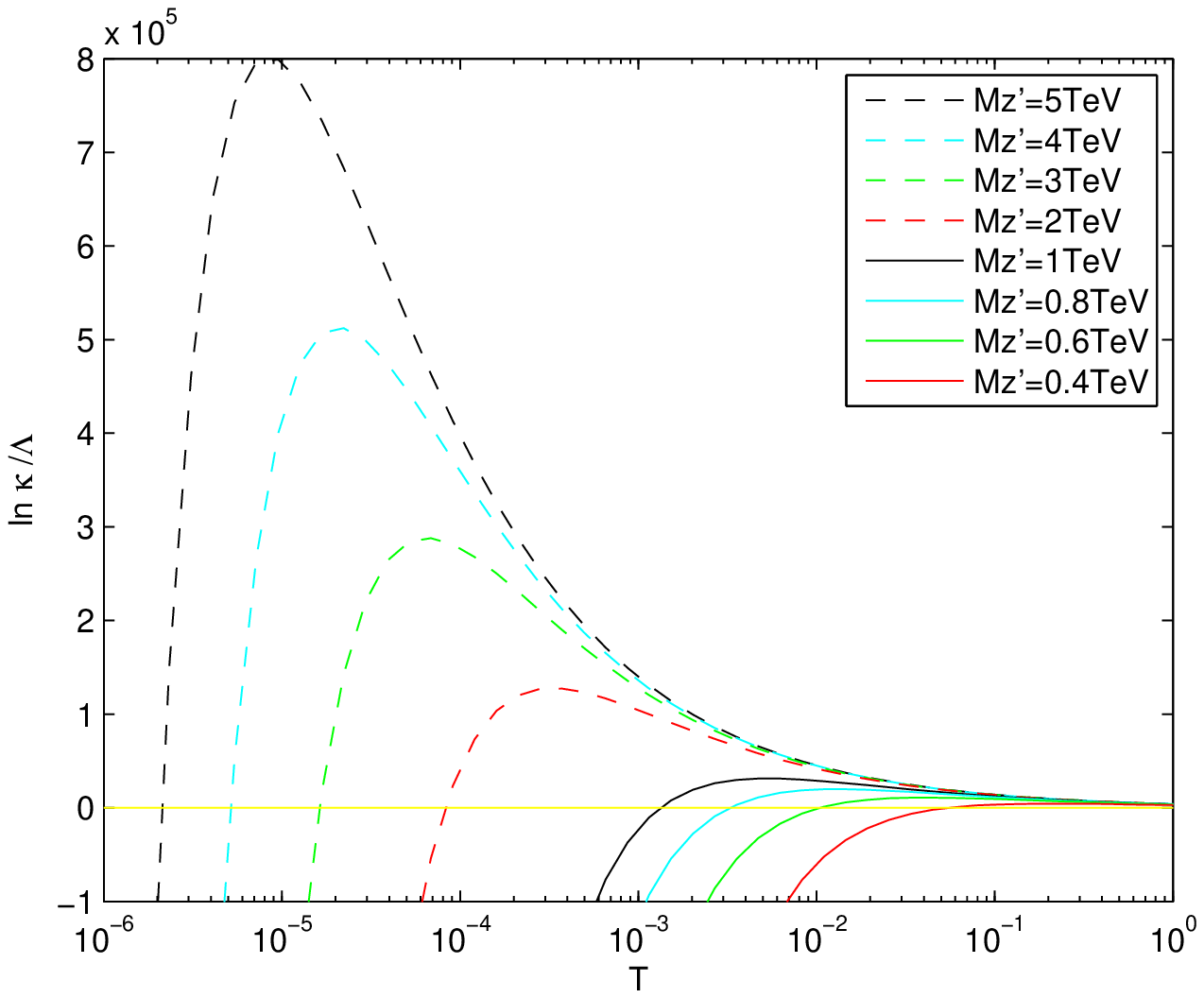}
\end{minipage}
\caption{ Upper bound of $Z'$ mass in unit of TeV as a function of
the $T$ parameter and $\kappa/\Lambda$.} \label{fig-cutoff2}
\hspace*{-4.5cm}\begin{minipage}[t]{5cm}
    \includegraphics[scale=0.6]{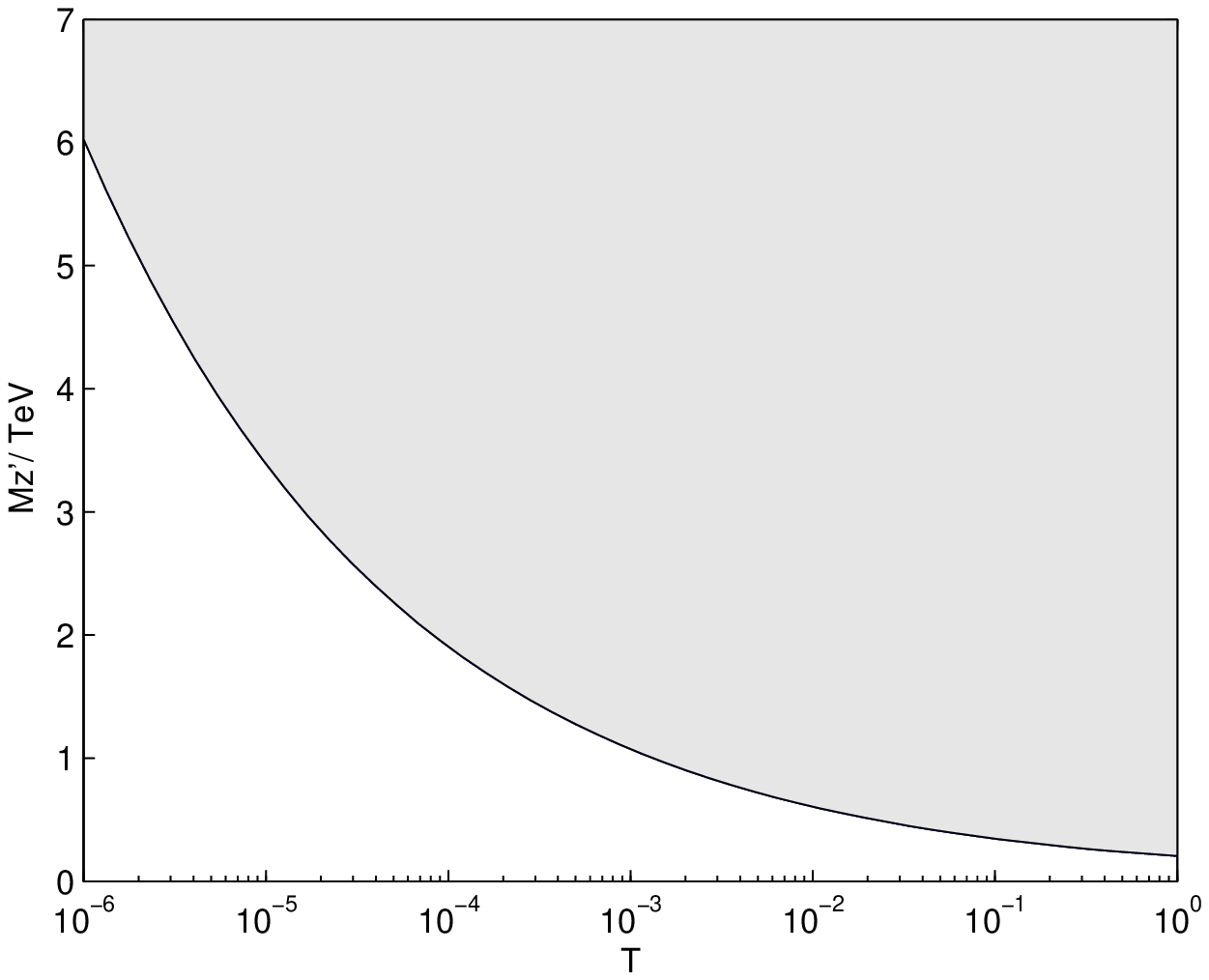}
\end{minipage}
\end{figure}

\begin{figure}[t]
\caption{The S parameter as functions of $T$ and $M_{Z'}$ }
\label{fig-S} \hspace*{-4.5cm}\begin{minipage}[t]{5cm}
    \includegraphics[scale=0.6]{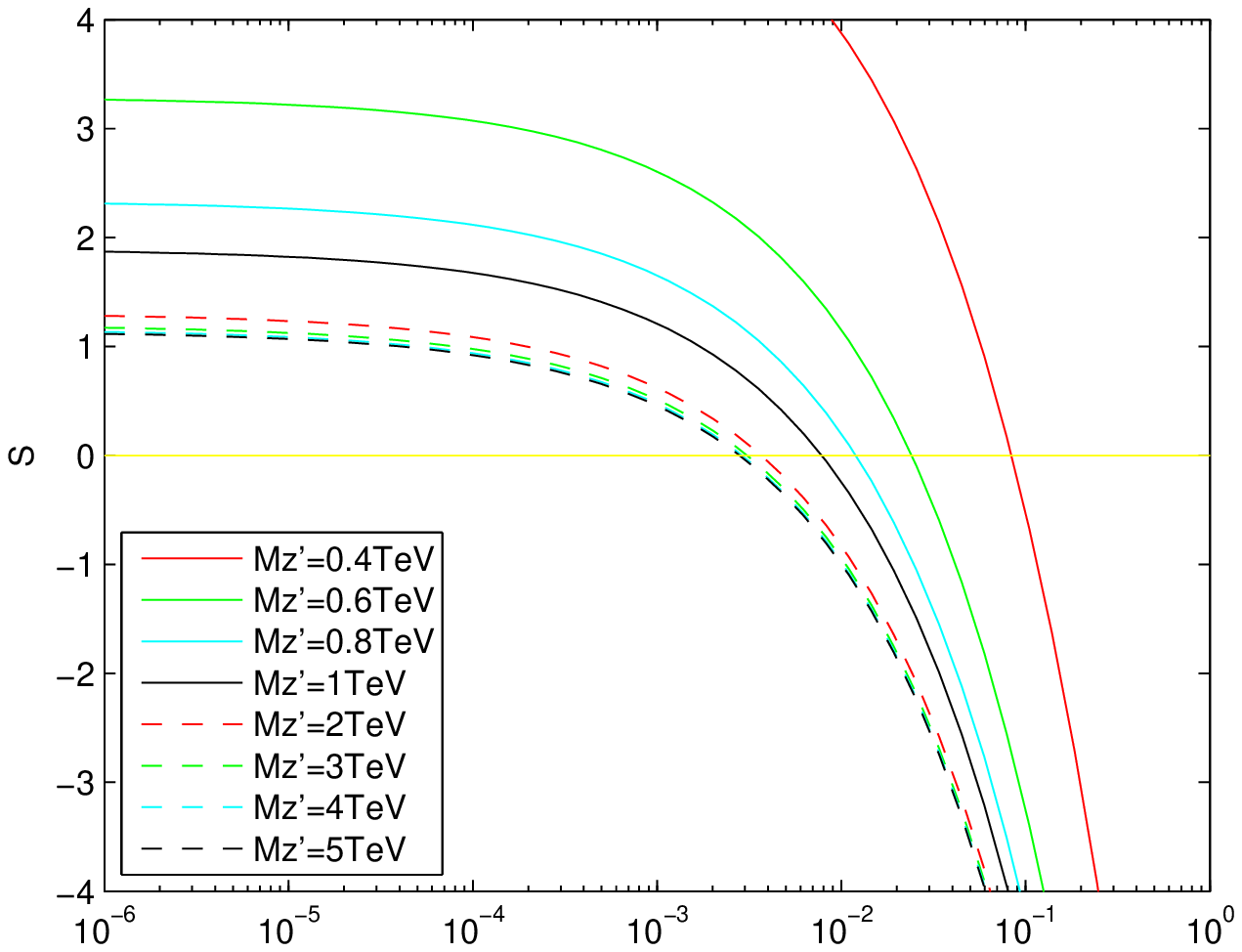}
\end{minipage}
\end{figure}

\begin{figure}[t]
\caption{$\alpha_1$ and $\alpha_2$ as functions of $T$ }
\label{fig-alpha12} \hspace*{-4.5cm}\begin{minipage}[t]{5cm}
    \includegraphics[scale=0.6]{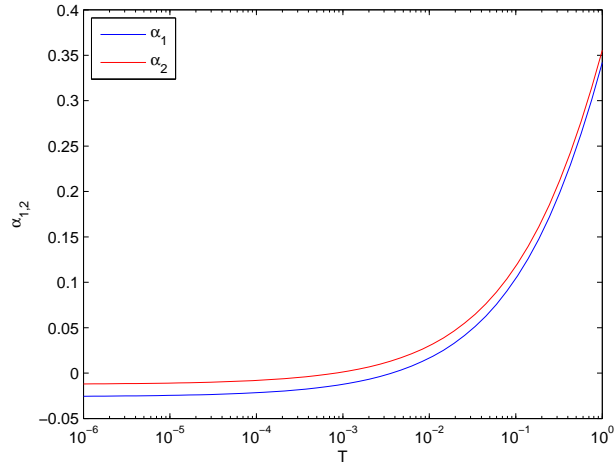}
\end{minipage}
\end{figure}

\begin{figure}[t]
\caption{$\alpha_3$, $\alpha_4$ and $\alpha_7$ as functions of $T$ }
\label{fig-alpha347} \hspace*{-4.5cm}\begin{minipage}[t]{5cm}
    \includegraphics[scale=0.6]{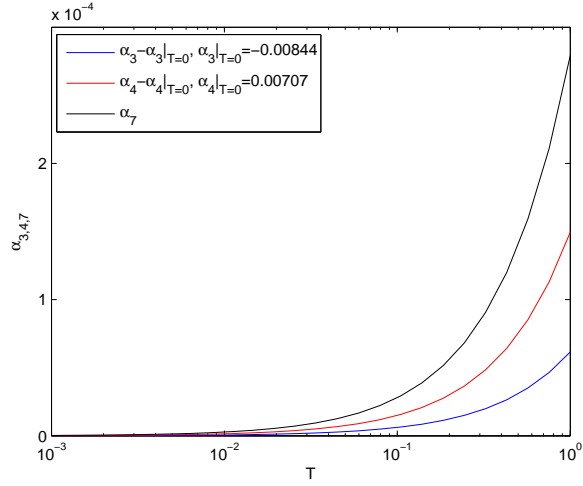}
\end{minipage}
\end{figure}

\begin{figure}[t]
\caption{$\alpha_5$,$\alpha_6$,$\alpha_8$ and $\alpha_9$ as
functions of $T$ } \label{fig-alpha5689}
\hspace*{-4.5cm}\begin{minipage}[t]{5cm}
    \includegraphics[scale=0.6]{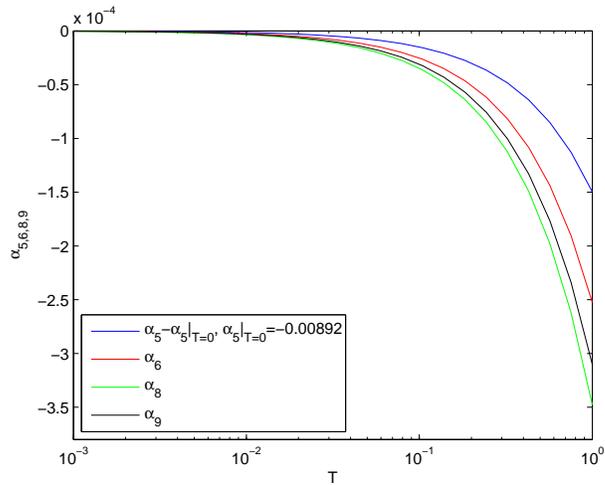}
\end{minipage}
\end{figure}

\begin{figure}[t]
\caption{$\alpha_{10}$ as a function of $T$ } \label{fig-alpha10}
\hspace*{-4.5cm}\begin{minipage}[t]{5cm}
    \includegraphics[scale=0.6]{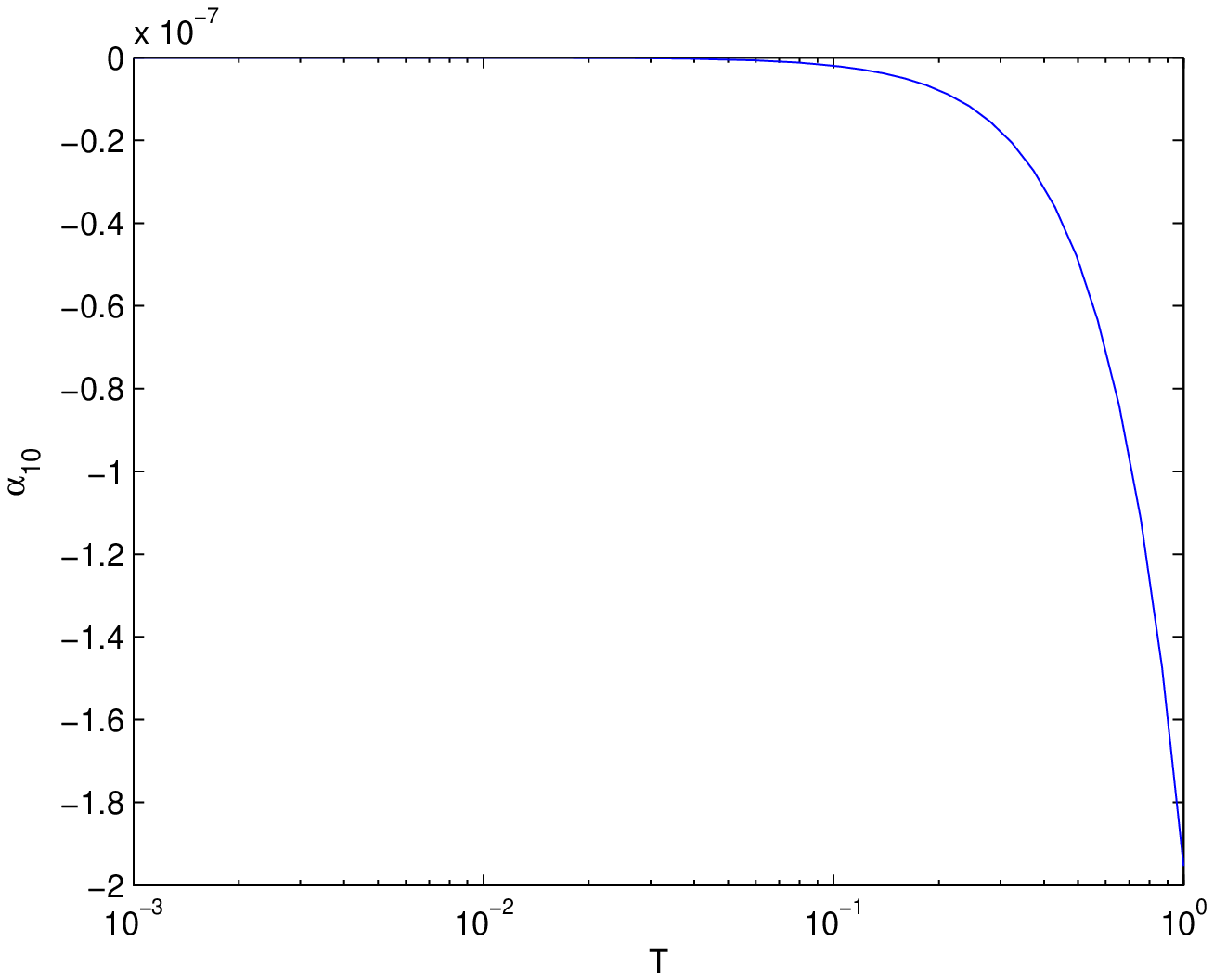}
\end{minipage}
\end{figure}

 The features of these results which are the same as those in K.Lane's
 model are:
\begin{enumerate}
\item The contributions to the $p^4$ order coefficients are divided into two parts:
the three doublets technicolor model contribution and the $Z'$
contribution.
\item All corrections from the $Z'$ particle are at
least proportional to $\beta_1$ which vanish if the mixing disappear
by $\theta'=0$.
\item Since $L_{10}^\mathrm{1D}<0$, combining with positive $\beta_1$, (\ref{alpha}) then tells us $\alpha_8$ is
negative. Then $U=-16\pi\alpha_8$ coefficient given in
Ref.\cite{EWCL} is always positive in present model.
\end{enumerate}
Since $\alpha_1$ and $\alpha_2$ depend on $\gamma$ which from
(\ref{gammaDef}) further rely on an extra parameter $\mathcal{K}$.
We can combine (\ref{f2beta1}),(\ref{cZ'def}) and (\ref{MZdef})
together to fix $\mathcal{K}$,
\begin{eqnarray}
\frac{3(F_0^{1D})^2g_1^2\cot^2\theta'}{8\beta_1M_{Z'}^2}&=&1+3\mathcal{K}g_1^{2}\cot^2\theta'
+\frac{3}{2}\mathcal{K}_{2}^{1D,\Sigma\neq0}g_1^2\cot^2\theta'+\frac{3}{2}\mathcal{K}_{13}^{1D,\Sigma\neq0}g_1^2
\cot^2\theta'\;.
\end{eqnarray}
Once $\mathcal{K}$ is fixed, with the help of (\ref{kappaDef}), we
can determine the ratio of infrared cutoff $\kappa$ and ultraviolet
cutoff $\Lambda$, in Fig.\ref{fig-cutoff1}, we draw the
$\kappa/\Lambda$ as functions of $T$ and $M_{Z'}$, we  find natural
criteria $\Lambda>\kappa$ offers stringent constraints on the
allowed region for $T$ and $M_{Z'}$ that present theory prefers
small $T$ parameter. The upper bound for $Z'$ mass increases as the
value of $T$ decrease, for example, upper bound is below 1TeV for
$T$ being order of $10^{-3}$ and 2-3TeV for $T$ being order of
$10^{-5}$. In Fig.\ref{fig-cutoff2}, we draw $Z'$ mass as a function
of $T$ parameter and the gray region is the forbidden zone where
 $\kappa\geq\Lambda$.
Not like K.Lane's model discussed in Ref.\cite{JunYi09} where we
have the upper bound of $Z'$ mass 400GeV, now this upper bound is
pushed higher as long as we have a very small $T$ parameter.
Considering that Ref.\cite{Sekhar} already gives lower bound of
$M_{Z'}=2.08$TeV, from Fig.\ref{fig-cutoff2} we find it corresponds
to $T<7.09\times 10^{-5}$. With this constraints on $M_{Z'}$, in
Fig.\ref{fig-S} we further draw the S parameter in terms of $T$ and
$M_{Z'}$. From this graph, we find that the S parameter in the
region of $T<7.09\times 10^{-5}$ and $M_{Z'}>2$TeV is still at order
of $+1$ which implies present model is still not fully matching with
the experiment data. Compared to previous result for K.Lane's
natural TC2 model with nonuniversal hypercharge assignments, we find
that the value of the S parameter does decrease due to the universal
hypercharge. For example, $S\approx 1.1$ at $T=10^{-2}$ for K.Lane's
model, while $S\approx 0$ at $T=10^{-2}$ for present model, this is
compatible with result obtained in Ref.\cite{Sekhar}, but for more
smaller T parameter, $S$ increases and finally for $M_{Z'}$ at
2-3TeV, $S$ is still at order of $+1$. Finally for completion of our
discussion, we depict all nonzero coefficients $\alpha_i$.
Fig.\ref{fig-alpha12} is the graph for $\alpha_1$ and $\alpha_2$,
Fig.\ref{fig-alpha347} is for $\alpha_3$, $\alpha_4$ and $\alpha_7$,
Fig.\ref{fig-alpha5689} is for $\alpha_5$, $\alpha_6$, $\alpha_9$
and $\alpha_8$, Fig.\ref{fig-alpha10} is for $\alpha_{10}$. In all
these diagrams, we find that the curves are not sensitive to
$M_{Z'}$ when $M_{Z'}>1-2$TeV, therefore we do not label the
$M_{Z'}$ on the graph. For Fig.\ref{fig-alpha347},
Fig.\ref{fig-alpha5689} and Fig.\ref{fig-alpha10}, the T axis starts
from $10^{-3}$ instead of $10^{-6}$, since below $T=10^{-3}$, all
curves approach to zero.

To summarize, we apply the formulation developed in
Ref.\cite{HongHao08} to a hypercharge-universal topcolor model,
compute all the coefficients of the bosonic part of EWCL up to the
order of $p^4$. We find that the universal hypercharge does improve
the model from the original nonuniversal hypercharge assignments by
allowing a larger $Z'$ mass resulting in a very small T parameter,
but the S parameter is still kept at order of $+1$.

\section*{Acknowledgments}

This work was  supported by National  Science Foundation of China
(NSFC) under Grant No. 10875065.




\begin{thebibliography}{1}\label{biblio}

\bibitem{Hill95}
C.T.Hill, Phys.Lett.B {\bf 345}, 483(1995)

\bibitem{Lane95}
K.Lane and E.Eichten, Phys.Lett. B {\bf 352}, 382(1995)

\bibitem{Sekhar}
F.Braam, M.Flossdorf, R.S.Chivukula, S.D.Chiara and E.H.Simmons,
Phys. Rev. D {\bf 77}, 055005(2008)

\bibitem{HongHao08}
H.H.Zhang, S.Z.Jiang, J.Y.Lang and Q.Wang, Phys. Rev. D. {\bf 77},
055003(2008)

\bibitem{JunYi09}
J.Y.Lang, S.Z.Jiang and Q.Wang, Phys. Rev. D. {\bf 79}, 015002(2009)

\bibitem{EWCL}
 T.Appelquist and G-H. Wu, Phys. Rev. {\bf D48},
3235(1993); {\bf D51}, 240(1995)


\end{thebibliography}
\end{document}